\documentstyle[11pt,newpasp_cv02,twoside,epsf]{article}
\def\edcomment#1{\iffalse\marginpar{\raggedright\sl#1\/}\else\relax\fi}
\reversemarginpar

\begin{document}
\title{Ab initio Simulations of Accretion Disc Boundary Layers} 

\author{Philip J. Armitage} 

\affil{University of St Andrews, School of Physics and Astronomy, 
	North Haugh, Fife KY16 9SS, UK}

\begin{abstract}
I discuss the results of simplified three dimensional 
magnetohydrodynamic simulations of the boundary layer 
between a disc and a non-rotating, unmagnetized star. 
Strong magnetic fields, possibly approaching equipartition 
with the thermal energy, occur in the boundary layer 
due to the shearing of disc-generated fields. The 
mean boundary layer magnetic field, which is highly variable 
on an orbital timescale, is estimated to  
exceed $\sim$50~kG for a 
CV with an accretion rate of $10^{-9} \ M_\odot {\rm yr}^{-1}$.
However, these fields do not drive efficient angular momentum 
transport within the boundary layer. As a consequence 
the radial velocity in the boundary layer is low, 
and the density high. 
\end{abstract}

\section{Introduction}
An accretion flow onto a weakly magnetic star must 
make a transition between the angular velocity in 
the disc, which is usually close to Keplerian, and 
the stellar angular velocity, which is typically much smaller. 
The boundary layer across which this transition 
occurs can be important both energetically, releasing 
up to half the bolometric luminosity of the accretion 
flow, and as a site of variability, in accretion 
onto protostars, white dwarfs and neutron stars.

\begin{figure}[t]
\plotone{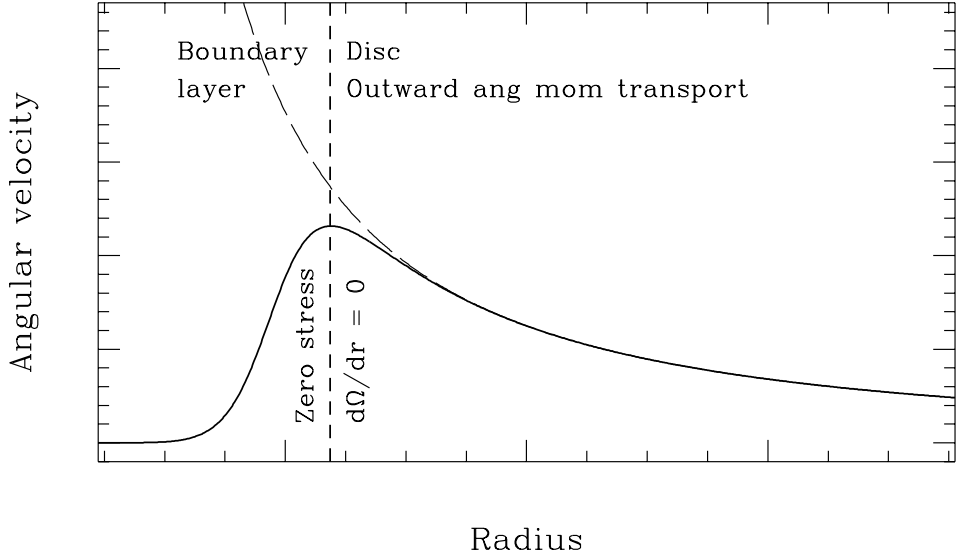}
\vspace{-2.2truein}
\caption{Illustration of the boundary layer structure for a  
	geometrically thin accretion flow in which angular momentum 
	transport can be described as a viscosity. In the disc, outward 
	angular momentum transport leads to mass inflow, and the angular 
	velocity is close to Keplerian (dashed curve). Close to 
	the star, the angular velocity reaches a maximum. There 
	is no viscous stress in the disc at this radius. At 
	smaller radii, inward angular momentum transport within the 
	boundary layer determines the boundary layer structure.}
\end{figure}

The aim of this contribution is to present the results 
of preliminary magnetohydrodynamic (MHD) simulations of 
disc boundary layers (Armitage 2002)\footnote{See also animations  
at {\tt http://star-www.st-and.ac.uk/$\sim$pja3/movies.html}}.
First though, it is useful to 
summarize the basics of boundary layer theory (e.g. 
Frank, King, \& Raine 1992).
To do so, let us forget magnetic fields for the time 
being, and assume that angular momentum transport in 
the accretion flow can be described as a viscosity, which   
depends upon the local shear  
in the angular velocity $\Omega$. In the disc, 
${\rm d}\Omega / {\rm d}r <0$, and angular momentum is 
transported outwards. If the star is slowly rotating, 
the inflowing gas inevitably 
reaches a maximum angular velocity $\Omega_{\rm max}$, as 
illustrated in Figure~1. At the radius $r_{\rm max}$ of this maximum, 
there is zero torque, so that the rate of accretion of angular momentum 
is simply $r_{\rm max}^2 \Omega_{\rm max} \dot{M}$.
Inside this radius, we reach 
the boundary layer region, where ${\rm d}\Omega / {\rm d}r > 0$. 
Angular momentum transport here is inward, and acts to mix
the accreted angular momentum into the star.

A somewhat more quantitative picture of the structure of the boundary layer 
follows from considering the radial component of the momentum 
equation. In a steady state, and ignoring viscous and magnetic 
terms, 
\begin{displaymath}
 v_r { {\partial v_r} \over {\partial r} } - { v_\phi^2 \over r }
 + {1 \over \rho} { {\partial P} \over {\partial r} } + 
 { {GM} \over r^2 } = 0,
\end{displaymath} 
where the symbols have their conventional meanings. In a geometrically 
thin disc the centrifugal term $v_\phi^2 / r$ accurately balances the 
gravitational force $GM / r^2$, but this is not 
the case in the boundary layer. If the boundary layer has 
radial width $\delta r_{\rm BL}$, there are evidently two 
possibilities:
\begin{itemize}
\item
Pressure forces balance gravity. Writing $P=\rho c_s^2$, where 
$c_s$ is the sound speed, we have $\rho^{-1} {\partial P} / 
{\partial r} \sim c_s^2 / \delta r_{\rm BL}$. If the boundary 
layer, like the disc, is to be geometrically thin, then 
$c_s \ll v_K$, the Keplerian orbital velocity. Pressure gradients 
can only balance gravity if $\delta r_{\rm BL} \ll r_*$.
\item
The $v_r  {\partial v_r} / {\partial r} \sim v_r^2 / \delta r_{\rm BL}$ 
term balances gravity. Causality demands that the radial velocity be 
smaller than the sound speed (Pringle 1977), so $v_r^2 \ll v_K^2$ 
and again we conclude that $\delta r_{\rm BL} \ll r_*$.
\end{itemize}
This analysis then suggests that the boundary layer will be 
radially narrow, {\em unless} the temperature becomes high 
enough to raise the sound speed to of the order of the Keplerian 
orbital velocity. Real calculations show that radially narrow boundary 
layers are expected in cataclysmic variables (Popham \& 
Narayan 1995), whereas broader boundary layers are possible 
in rapidly accreting pre-main-sequence stars such as 
FU Orionis objects (Popham 1996; Kley \& Lin 1999), and 
around neutron stars (Popham \& Sunyaev 2001). A broad 
dynamical boundary layer has the interesting property 
that it allows accretion to occur without spinning up 
the star, even at stellar rotation rates well below 
break up (Popham \& Narayan 1991).

To go beyond these general considerations, it is necessary to 
know something about the angular momentum transport in the 
boundary layer, since this will determine the radial velocity and 
density structure. The traditional choice in the {\em disc} is 
to express the viscosity using the Shakura \& Sunyaev (1973) $\alpha$ 
prescription,
\begin{displaymath}
 \nu = \alpha c_s h
\end{displaymath} 
where $h$ is the disc scale height and 
$\alpha$ is a constant. Unfortunately this proves 
disastrous if extended into the boundary layer, since it can 
lead to unphysical supersonic radial velocities. To remedy this, 
alternatives to the $\alpha$ prescription have been proposed, 
which have the effect of reducing $\nu$ in the boundary layer. 
A simple example is to replace the vertical scale height $h$
in the above equation with the (smaller) radial pressure 
scale height $h_r$, though more elaborate schemes have also 
been devised (Lynden-Bell \& Pringle 1974; 
Papaloizou \& Stanley 1986; Popham \& Narayan 1992).

\section{The role of magnetic fields}

How might magnetic fields affect this picture? One 
possibility is that magnetic fields, generated in the 
disc by magnetorotational instabilities (Balbus \& Hawley 1991), 
could be amplified by shear in the boundary layer (Pringle 1989).
The basic agument is that the rate of shearing in the 
boundary layer exceeds that in the disc by a factor 
$\sim r_* / \delta r_{\rm BL}$, which is large, perhaps 
10-100. Unless the loss processes for the magnetic field 
(for example Parker instability, or reconnection) 
are similarly accelerated, the result will be a strong, 
predominantly toroidal field. Pringle (1989) estimated 
that the resultant Alfv\'en speed in the boundary 
layer could be of the order of,
\begin{displaymath}
 v_A \sim \left( { h \over r} \right)^{1/2} v_K,
\end{displaymath} 
though a contemporary estimate would be somewhat lower 
since MHD turbulence in discs leads to poloidal fields 
that are substantially smaller than the toroidal component.
Nonetheless, substantial field amplification appears 
possible. I note that analagous processes have been considered 
in the context of black hole accretion (Krolik 1999), 
where the shear arises in the dynamically unstable 
region interior to the marginally stable orbit.

Such field amplification does not depend upon the existence 
of magnetic instabilities or dynamo processes in the 
boundary layer, and is therefore a rather robust prediction.
The origin and magnitude of angular 
momentum transport in the boundary layer is, however, an interesting question 
in its own right. The identification of magnetorotational 
instabilities (MRI) as the key ingredient for disc angular 
momentum transport leaves the situation in the boundary 
layer unclear. The condition for the MRI to operate is 
that,
\begin{displaymath} 
 { { {\rm d} \Omega^2 } \over { {\rm d} r } } < 0,
\end{displaymath} 
which is crucially less stringent than the Rayleigh 
criteria for hydrodynamic instability, 
${\rm d} (r^2 \Omega) / {\rm d} r < 0$. As a result, 
discs are always MRI unstable. Fields in the boundary 
layer, however, would be stable by this criteria.
This has two implications. First, it suggests that 
the viscosity in the boundary layer might well be 
expected to be reduced as compared to the disc. 
This is promising, as we have already noted that 
existing models {\em require} such a reduction in the 
efficiency of angular momentum transport. Second, 
it means that fundamentally different processes 
must be responsible for angular momentum transport 
in the boundary layer. This is not so good, as 
it undermines the various prescriptions for 
boundary layer viscosity and suggests that 
existing quantitative models for the boundary layer 
are likely to be of limited validity.

\section{Numerical simulations}

The ZEUS code (Stone \& Norman 1992a, 1992b, Norman 2000) is used 
to solve the equations of ideal MHD. ZEUS is an explicit, Eulerian 
MHD code, which uses an artificial viscosity to capture shocks.
The simulations model the inner disc, boundary layer, and hydrostatic 
envelope of the star, and are fully three-dimensional. 
Simplifications are made by assuming an 
isothermal equation of state, and by ignoring the vertical 
component of gravity in the disc structure. Simulating an unstratified disc  
reduces the computational burden, and should 
be a reasonable approximation close to the equatorial plane.
The simulation geometry is cylindrical, $(z,r,\phi$), using uniform gridding 
in the vertical and azimuthal directions, and a grid in the 
radial direction which concentrates resolution close to the boundary 
layer. The highest resolution calculation discussed here 
used 480 radial, 90 azimuthal, and 60 vertical grid cells.

The initial conditions for the simulation comprise a static, 
non-rotating and unmagnetized atmosphere, surrounded by a  
Keplerian disc with an approximately gaussian density profile.
The disc is initially seeded with a weak vertical magnetic field 
to initiate magnetorotational instabilities.
A boundary layer then forms automatically once the MRI is active 
and mass in the disc is flowing inwards.
The boundary conditions are periodic in $z$ and $\phi$, reflecting 
at $r = r_{\rm in}$ ($v_r = B_r = 0$), and set to outflow at 
$r = r_{\rm out}$.

\section{Results}

\begin{figure}[t]
\plotfiddle{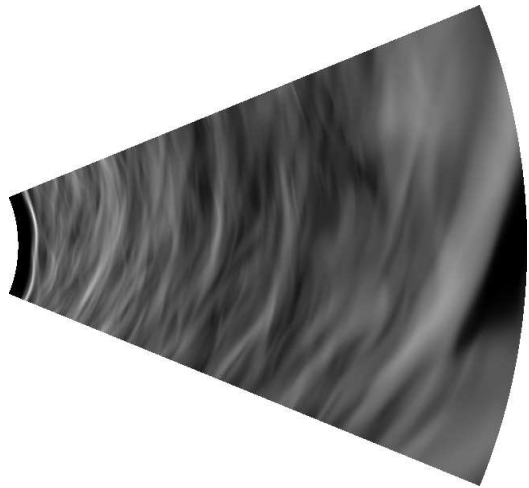}{2.5truein}{0}{75}{75}{-225}{-195}
\vspace{-0.2truein}
\caption{Map of the magnetic field strength as a fraction 
	of the thermal energy in the simulation, integrated 
	vertically. The boundary layer is the narrow bright 
	stripe of field near the inner edge of the disc. 
	The radial grid in the simulation is much finer at 
	small $r$ to concentrate resolution in the boundary layer 
	region.}
\end{figure}

Figure 2 shows the strength of the 
magnetic fields generated towards the end of the simulation, when 
the MRI in the disc has reached a saturated and quasi-steady state. 
The azimuthal domain is $45^\circ$ 
across, and the grid extends from an inner edge at $r_{\rm in}=1$ 
(in arbitrary units) to $r_{\rm out}=5$. The innermost region 
is occupied by high density, pressure supported gas, which is 
initially unmagnetized and remains so during the course of the 
run. Outside this is a narrow, strongly magnetized boundary 
layer, which appears as a bright stripe in the magnetic field 
map. This region is resolved across 20-30 grid points in the 
highest resolution run so far completed.
An animation of the evolution shows that the strong 
field region in the boundary layer is highly variable, buckling 
and changing in strength on a dynamical timescale. At larger 
radii the gas in the accretion disc is in close to Keplerian 
rotation, and displays the complex pattern of magnetic fields 
seen in previous MHD simulations of disc turbulence.

\subsection{Magnetic field strength}

\begin{figure}[t]
\plottwo{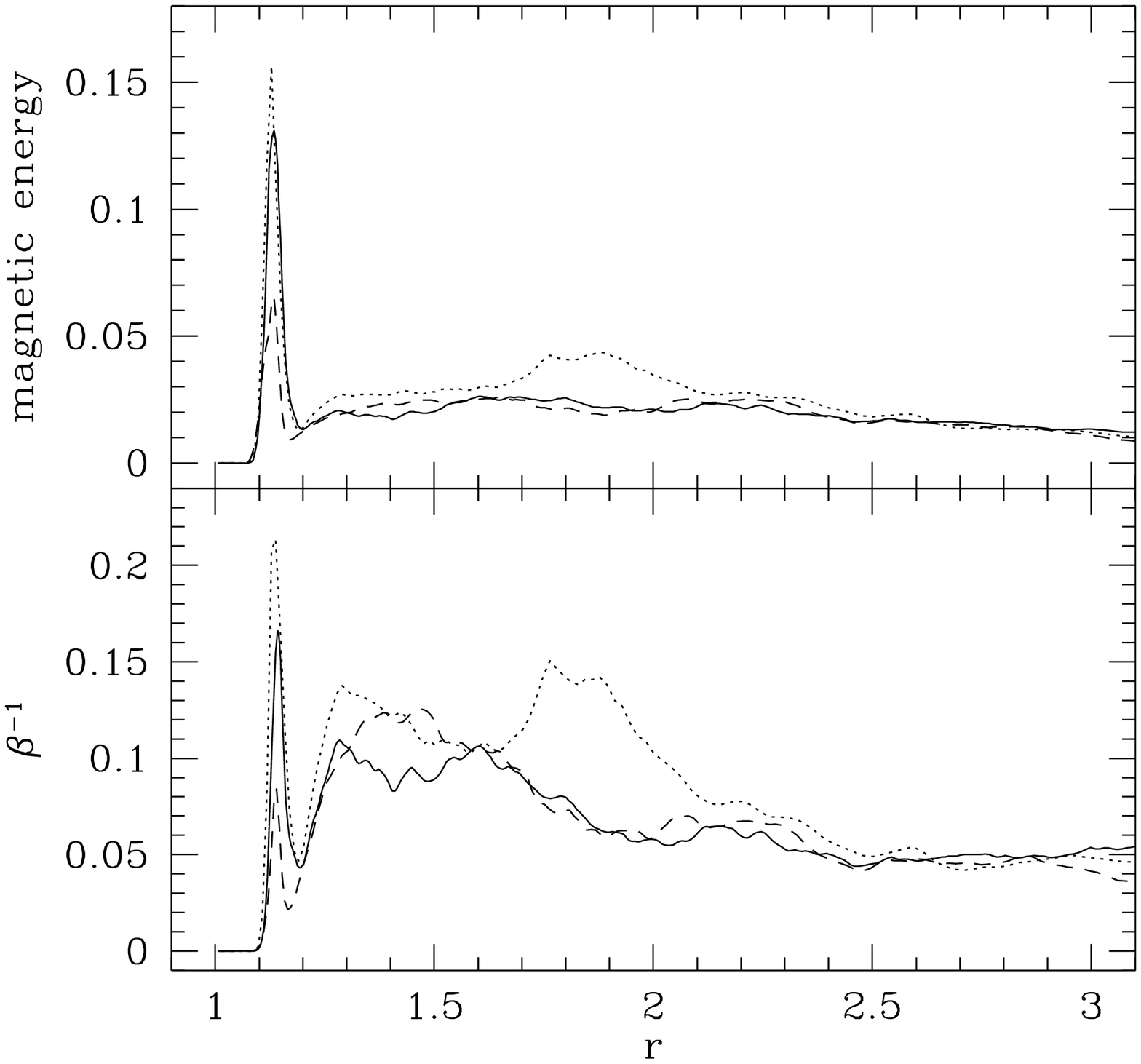}{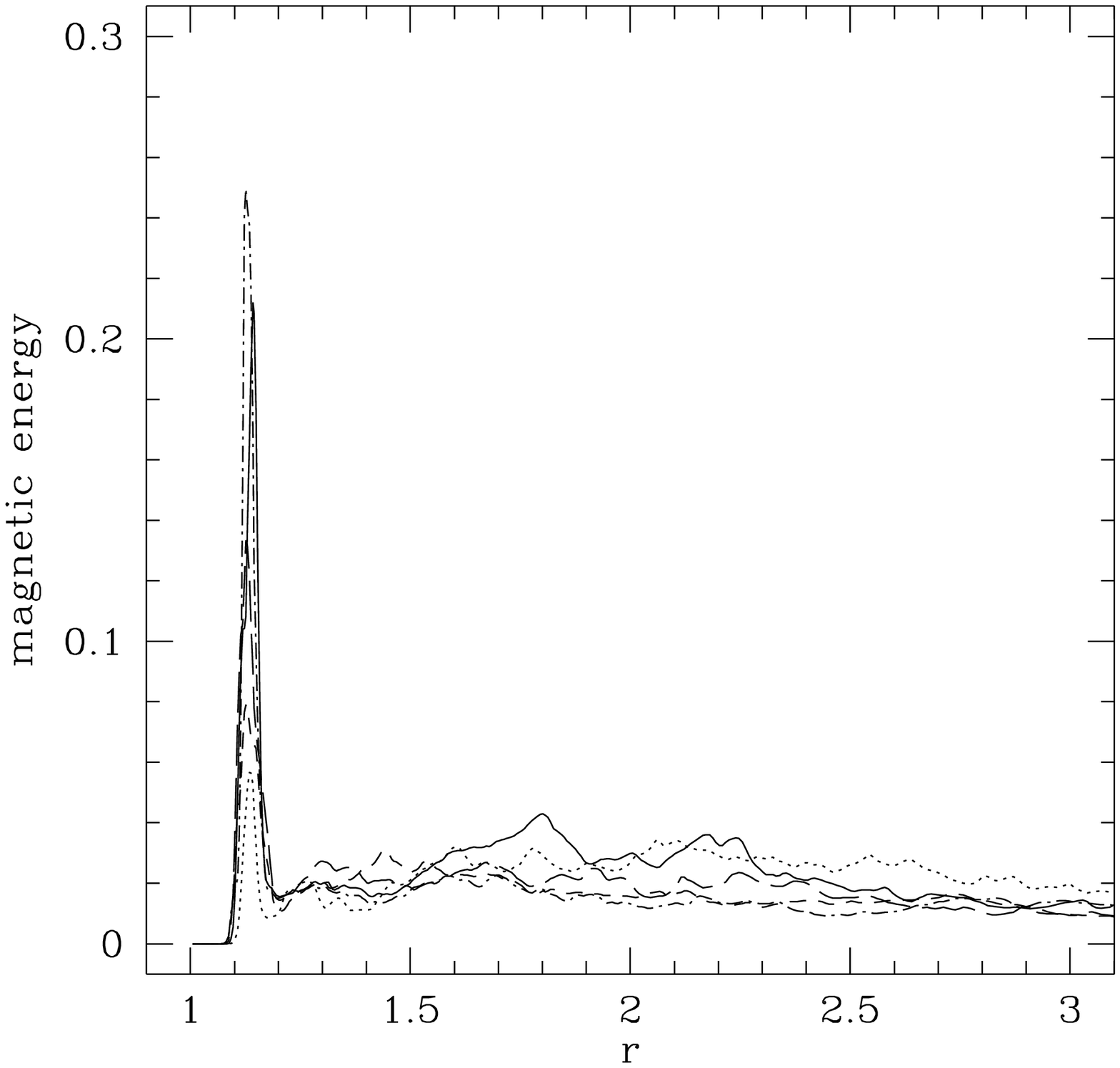}
\caption{Strength of the magnetic fields in the disc and boundary layer.
	The lefthand plot shows the mean (averaged over several independent 
	timeslices) magnetic energy density in three 
	simulations with different resolution, both in absolute 
	terms (upper panel) and as a fraction of the thermal energy 
	(lower panel). The righthand plot shows slices from one 
	simulation at different times. There are large fluctuations 
	in the boundary layer field strength.}
\end{figure}

Figure 3 shows the time-averaged strength of the magnetic fields generated in the 
simulation, both in terms of the absolute magnetic field energy density 
(in arbitrary code units), and as a fraction $\beta^{-1}$ of the thermal 
energy in the disc. The expected amplification of disc magnetic fields 
by shear in the boundary layer is clearly seen, with the magnetic 
field energy in the boundary layer exceeding that in the disc 
immediately outside by around an order of magnitude. The same 
conclusion applies -- albeit less strikingly -- if the magnetic 
energy is compared to the thermal energy. In the disc, the energy 
in the magnetic fields generated in the simulation is only around 
5\% of the thermal energy. In the boundary layer, this rises to 
around 15\%.

Figure 3 also shows how the magnetic fields vary with time. Fields 
in the boundary layer are highly variable -- even more so than in 
the disc. This makes it hard to judge, given the limited period 
of evolution simulated, whether the 
resolution is sufficient to have achieved numerical convergence 
in the boundary layer. Pending higher resolution runs, the 
field strengths quoted above are therefore probably lower limits.

\subsection{Angular momentum transport}

\begin{figure}[t]
\plottwo{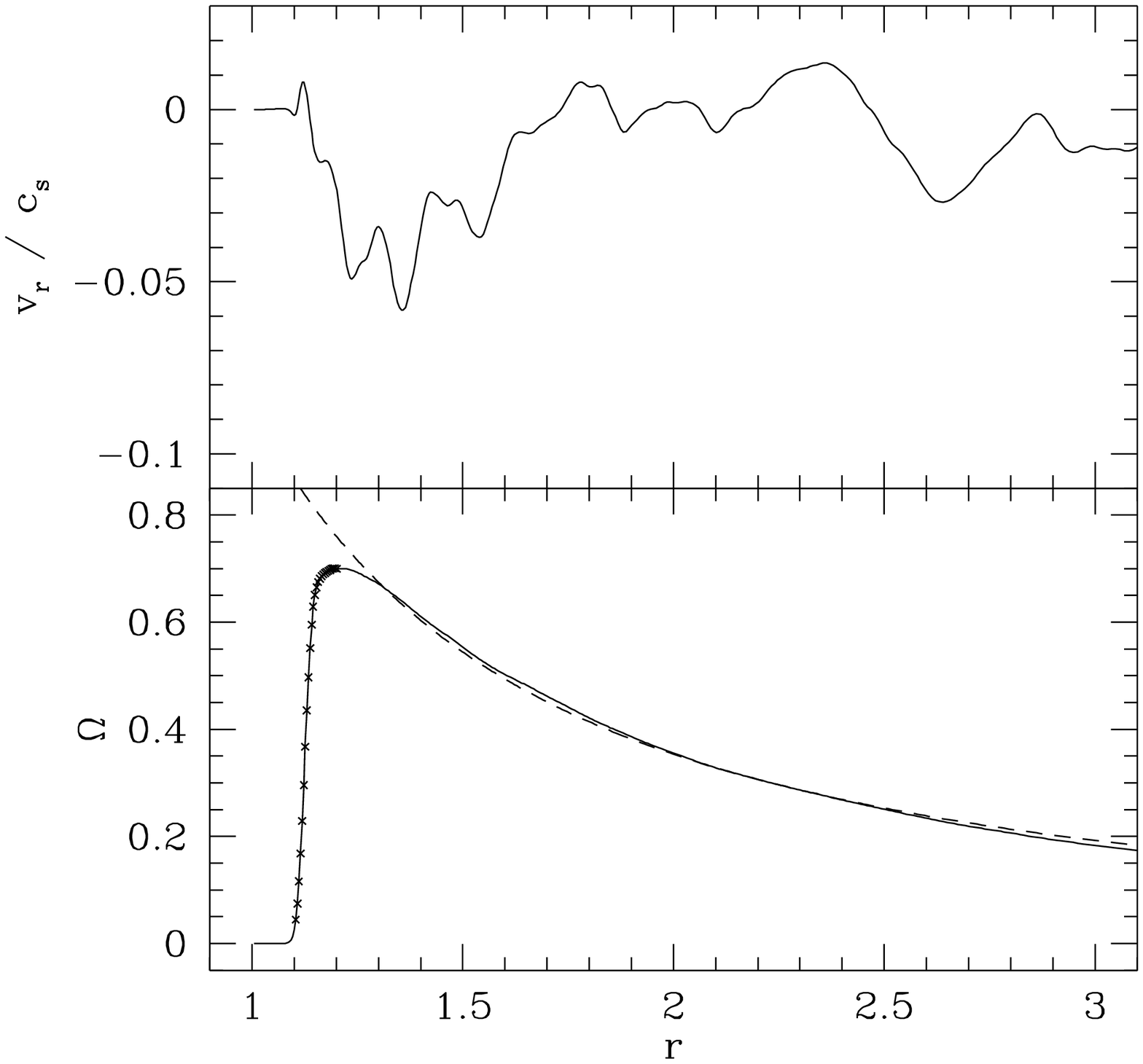}{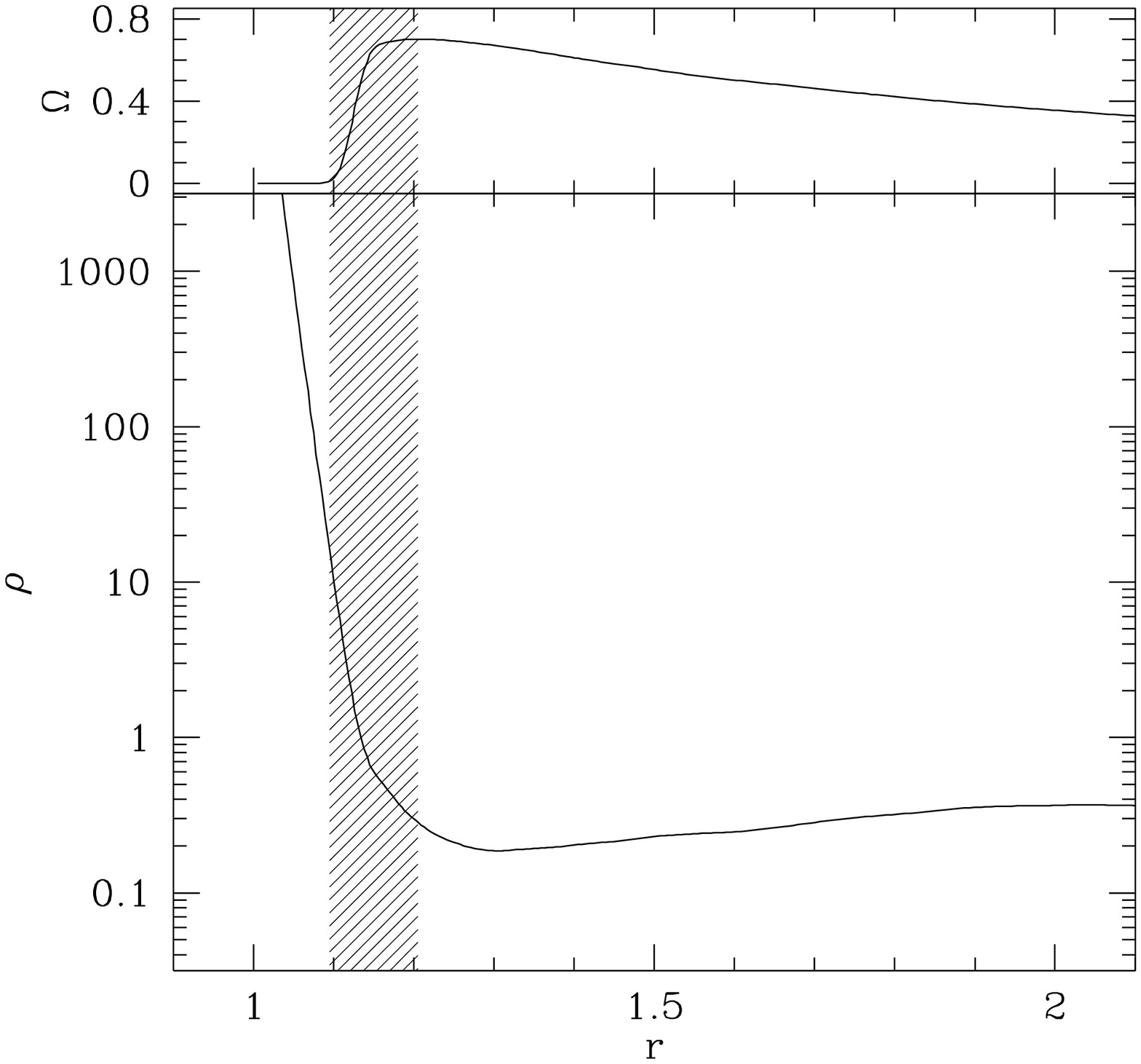}
\caption{Left panel: the mean radial and angular velocity in the 
	simulation. The radial velocity in the boundary layer is 
	low, so the density (right panel) in the boundary layer 
	is relatively high.}
\end{figure}

Figure 4 shows the angular and radial velocities obtained 
in the simulations, again averaged over several timeslices. 
The radial velocity is everywhere highly subsonic, and 
is largest in the disc just outside the boundary layer. 
The small radial velocity in the boundary layer occurs  
because angular momentum transport in this region is 
inefficient, and as a result the boundary layer occupies 
relatively high density gas (right panel of the Figure) 
adjacent to the stellar envelope. Analysis of the 
magnetic stress as a function of radius suggests that 
the stress acts like a viscosity in so far as it vanishes 
very close to the radius where ${\rm d} \Omega / {\rm d}r = 0$. 
In the boundary layer itself, the magnetic stress 
is very small.

This lack of angular momentum transport in the 
boundary layer implies a different boundary layer 
structure as compared to one-dimensional models. 
For the rather thin boundary layer simulated here, 
however, the differences may not be large, since 
all existing boundary layer viscosity prescriptions imply a 
sharp reduction in the efficiency of angular momentum 
transport in a narrow boundary layer. More substantial 
differences may occur for boundary layers of greater 
radial extent.

\subsection{Scaling to observed systems}

The current simulations are too rudimentary to be able to provide 
a self-contained measure of the boundary layer field strength 
in real systems. However, we can make an order of magnitude estimate 
by making use of the scaling relations for the central density $\rho$
and temperature $T_c$ of $\alpha$ discs. For a Kramers' opacity 
($\kappa \propto \rho T_c^{-7/2}$) Frank, King \& Raine (1992) 
quote,
\begin{eqnarray}
 \rho & = & 3.1 \times 10^{-8} \alpha^{-{7 \over 10}} 
 \left( { \dot{M} \over {10^{16} \ {\rm gs}^{-1} } } \right)^{11 \over 20} 
 \left( { M \over M_\odot } \right)^{5 \over 8} 
 \left( { r \over {10^{10} \ {\rm cm} } } \right)^{-{15 \over 8}} 
 f^{11 \over 5} \ {\rm gcm}^{-3} \nonumber \\
  T_c & = & 1.4 \times 10^{4} \alpha^{-{1 \over 5}} 
 \left( { \dot{M} \over {10^{16} \ {\rm gs}^{-1} } } \right)^{3 \over 10} 
 \left( { M \over M_\odot } \right)^{1 \over 4} 
 \left( { r \over {10^{10} \ {\rm cm} } } \right)^{-{3 \over 4}} 
 f^{6 \over 5} \ {\rm K} \nonumber \\
 f & = & \left[ 1 - \left( {r_* \over r} \right)^{1 \over 2} \right]^{1 \over 4}. \nonumber
\end{eqnarray} 
Appropriate values in the inner regions of CV discs are a radius of 
$\approx 10^9 \ {\rm cm}$, a white dwarf mass of $0.6 M_\odot$, and 
an $\alpha$ parameter of $\sim 0.1$.

With these scaling relations in hand, we can estimate the gas pressure 
in the disc at $r=10^9 \ {\rm cm}$, near the boundary layer. The simulations 
can then be used to estimate first the magnetic energy in the disc, and from
that the magnetic energy in the boundary layer. The results suggest that the 
magnetic energy density in the bounday layer is roughly comparable
to the thermal energy in the inner disc, which implies,
\begin{displaymath}
 B_{\rm BL} \simeq 50 \left( {\dot{M} \over {10^{-9} \ M_\odot {\rm yr}^{-1}} } 
 \right)^{17/40} \ {\rm kG}.
\end{displaymath} 
Given the limitations of the simulations 
this figure is probably a lower limit to the 
true field strength, and is definitely subject to considerable 
uncertainty. It should be regarded as an order of magnitude 
estimate only.

\section{Summary}

This contribution has discussed three-dimensional 
MHD simulations of accretion disc boundary layers. 
The main results of the work so far are,
\begin{itemize}
\item
Angular momentum transport in the boundary layer is 
inefficient. This means that the radial velocity in the 
boundary layer is highly subsonic, and the density high.
Although this is qualitatively consistent with the modified 
viscosity prescriptions used in one-dimensional models 
of the boundary layer, such prescriptions have no physical 
basis, and results derived from them need to be treated with caution.
\item
Strong magnetic fields are generated in the boundary layer 
from shearing of disc fields. For weakly magnetized stars, 
the strongest magnetic fields will be in the boundary layer.
\item
The magnetic fields and boundary layer generally are highly 
variable.
\end{itemize}
Further simulations, incorporating some aspects of the 
thermal structure of the boundary layer, are needed to 
investigate these questions in more detail.

\acknowledgements{My thanks to Andrew Cumming, Kristen Menou 
and Gordon Ogilvie for stimulating discussions  
during the course of the meeting. Computations made 
use of the UK Astrophysical Fluids Facility.}

\end{document}